\documentclass[prl,showpacs,superscriptaddress,nofootinbib]{revtex4-2}
\usepackage{natbib}
\usepackage{graphicx}
\usepackage{xcolor}
\usepackage{easyReview}%
\usepackage{bm}
\usepackage{fleqn}
\usepackage{amssymb}
\usepackage{color}
\usepackage{amsmath}
\usepackage{amstext}
\usepackage{latexsym}
\usepackage{xcolor}

\begin{document}
\title{Single energy measurement Integral Fluctuation theorem \\ and non-projective measurements}

\author {Daniel Alonso and Antonia Ruiz García}
\affiliation{Departamento de F\'\i sica and IUdEA, \\ Universidad de La Laguna, La Laguna, Tenerife, Spain}

\begin{abstract}
We study a Jarzysnki type equality for work in systems that are monitored using non-projective unsharp measurements. The information acquired by the observer from the outcome $f$ of an energy measurement, and the subsequent conditioned normalized state $\hat \rho(t,f)$ evolved up to a final time $t$ are used to define work, as the difference between the final expectation value of the energy and the result $f$ of the measurement. The Jarzynski equality obtained depends on the coherences that the state develops during the process, the characteristics of the meter used to measure the energy and the noise it induces into the system. We analyze those contributions in some detail to unveil their role. We show that in very particular cases, but not in general, the effect of such noise gives a factor multiplying the result that would be obtained if projective measurements were used instead of non-projective ones. The unsharp character of the measurements used to monitor the energy of the system, which defines the resolution of the meter, leads to different scenarios of interest. In particular, if the distance between neighboring elements in the energy spectrum is much larger than the resolution of the meter, then a similar result to the projective measurement case is obtained, up to a multiplicative factor that depends on the meter. A more subtle situation arises in the opposite case in which measurements may be non-informative, i.e. they may not contribute to update the information about the system. In this case a correction to the relation obtained in the non-overlapping case appears. We analyze the conditions in which such a correction becomes negligible. We also study the coherences, in terms of the relative entropy of coherence developed by the evolved post-measurement state. We illustrate the results by analyzing a two-level system monitored by a simple meter.    
\end{abstract}

\date{\today}
\maketitle
\vskip0.5cm

\noindent

\section{Introduction}

Fluctuation Theorems have been one of the most relevant results of statistical physics in recent decades. The pioneering work of Evans, Morris and Searles \cite{evans1993probability,evans1994equilibrium}, Gallavotti, Cohen \cite{gallavotti1995dynamicalensembles,gallavotti1995dynamical}, Jarzynski \cite{jarzynski1997nonequilibrium,jarzynski1997equilibrium}, and Crooks \cite{crooks1999a} has been followed by a series of significant results that have highlighted their conceptual depth, as well as its relevance and applicability in many different fields  \cite{jarzynski2007comparison,jarzynski2011equalities,esposito2009nonequilibrium,campisi20111,seifert2012stochastic}. In particular, work fluctuations are one of the issues that have attracted more attention. The most widely used approach to study work fluctuations considers a two-point measurement scheme, in which work is defined as the difference between the outcomes of two projective energy measurements \cite{kurchan2000aquantum}. Specifically, given an initial state, a first projective energy measurement is made, afterwards the system evolves unitarily under the dynamics dictated by a time dependent Hamiltonian. Then, at a given final time, a second projective energy measurement is performed. Work is defined as the difference between the energies obtained in the two measurements. Although this scheme has been widely explored and has been very fruitful, there are still important open issues that need to be addressed. 

After the first energy measurement, the system is projected onto an eigenstate of the initial Hamiltonian. The unitary evolution that follows generally develop coherences in the eigenbasis of the final Hamiltonian. Such coherences disappear once the second energy measurement is performed. In this sense the two-point measurement scheme does not include some important quantum features. 

It has been shown two physically necessary properties of quantum work, namely, respecting the classical limit and obeying the first law of thermodynamics, cannot be simultaneously measured. Considering these results, schemes have been proposed to estimate fluctuating work using global measurements in which the backaction of the measurement can be reduced, and thus approximately describe coherent transformations \cite{acin2017nogo}.

Another aspect concerns the nature of the measurements performed. In general, if the energy measurement is made using a meter that has a finite resolution, the result obtained may not correspond to any of the eigenenergies of the measured Hamiltonian \cite{braginsky2003quantum}. Even so, the measurements may be sufficiently informative to conclude that the post-measurement state will be a given eigenstate of the Hamiltonian. In general, the meter will induce some noise into the system, and therefore work fluctuations will be affected \cite{watanabe2014generalized,watanabe2014generalized,chiara2015measuring}. Once the first measurement has been made, knowledge of the conditioned post-measurement state, as well as the unitary operator that dictates the subsequent time evolution provide sufficient information to obtain the expected value of the energy at the final time. In the case of projective measurements, the difference between such energy value and the result obtained in the first measurement has been used as a definition for work in \cite{deffner2016quantum,allahverdyan2005fluctuations,campisi2013quantum,sone2020quantum}. Remarkably, this definition for work makes possible to derive a modified Jarzynski quantum equality that evidences the role played by the projective measurements and their cost \cite{deffner2016quantum}. Notice that the possibility of obtaining work fluctuations from a single measurement was already pointed out in \cite{roncaglia2014work} and for open quantum systems and end-point measurement scheme has been proposed \cite{gherardini2021endpoint}. Also, several measurement protocols have been proposed to obtain the probality distribution of work done on a quantum system using an ancilla or quantum detector. The key point in these schemes is the system-detector interaction considered to couple the system energy operator with an observable of the detector. Then, the information about work statistics is encoded in the quantum state of the detector \cite{chiara2018ancilla}. More recently it has been shown that the quantum Jarynski equality arises as a particular case of a general quantum fluctuation theorem for the entropy production of an open quantum system coupled to multiple environments, not necessarily at equilibrium \cite{chiara2022entropy}.

In the limit of non-precise measurements, it is interesting to study how non-projective measurements affect work fluctuations and unveil the role of coherences. Also, the effect of the meter used to measure the energy, the relevance of the initial state, and its average energy in the quantum Jarzynski relation that would emerge, could be analyzed.


In this work, we begin by briefly introducing the two energy measurement approach mostly used in quantum mechanical analysis of work. In particular, we show the Jarsynski equality that arises when all energy measurements are projective.
We then focus on a more general setting in which the meter that monitors the system performs unsharp  measurements in which the outcomes are not reproducible. We show how these non-projective measurements can be described. Within this approach, we consider a definition of work based on a single non-projective energy measurement performed at a given time and the subsequent unitary evolution of the post-measurement state. We obtain the generalized Jarsynski equality that arises in this case, and which makes explicit the role of the measuring apparatus and coherences in work fluctuations. We perform a detailed analysis of the different effects that contribute to such equality. Specifically, we study the coherences, the noise induced by measurement and the information acquired by the observer from measurement outcomes. To illustrate the analysis we consider a two-level system monitored by a simple meter that performs unsharp measurements, characterized by a given conditional probability of outcomes. We study the different scenarios that arise depending on the measurement conditions, defined by the energy resolution of the meter in comparison to the energy spectrum of the system.

\subsection{Two-point measurement protocol}

The definition of work has been widely discussed in quantum mechanical analysis of work fluctuations \cite{campisi20111,manzano2018thermodynamics} as it is known that no hermitian operator can be assigned to work \cite{talkner2007c}. The most widely used approach is based on two energy measurements, the first one at the beginning of the process and the second one at the end. Work is defined as the difference of the two energy outcomes \cite{kurchan2000aquantum,tasaki2000jarzynski}. There are also some proposals based on a single measure of energy, whereby work can be obtained from a single final generalized energy measurement \cite{roncaglia2014work}. We now briefly introduce the two-point protocol to fix the notation.

We consider a system described by a time dependent Hamiltonian $\hat H(\lambda(t)):=\hat H(t)$, where $\lambda(t)$ defines the protocol corresponding to an external driving. We assume that at initial time $t=t_0$ the system is in thermal equilibrium with a thermal bath at $T$ temperature. Thus, it is in a Gibbs state given by $\hat \rho_{th}(t_0)={\cal Z}(t_0)^{-1} \exp{(-\beta \hat H(t_0))}$, with $\beta=(k_B T)^{-1}$ and ${\cal Z}(t_0)=\text{Tr }\exp{(-\beta \hat H(t_0))}$, where the trace is over the Hilbert space of the system. 

In the so called \emph{forward protocol}, at time $t=t^+_0$ it is performed a projective measurement of the energy of the system, given by $\hat H(t_0)=\sum_{\mu} \mu |\mu \rangle  \langle \mu|\equiv \sum_{\mu} \mu \,\hat \Pi_{\mu}$. If the result of the measurement is $\mu$, the post-measurement state becomes 
\begin{equation}
\hat \rho(t^+_0,\mu)= P_{\mu}^{-1} \hat \Pi_{\mu} \hat \rho_{th}(0) \hat \Pi_{\mu},
\label{eq:1}
\end{equation}  
with $P_{\mu}=\text{Tr } \hat \Pi_{\mu} \hat \rho_{th}(0)$ the probability of finding the energy $\mu$ as a result of the measurement. After this first energy measurement, the system evolves up to time $t=t_1$ according to the unitary evolution operator $\hat U(t_1,t_0)$, which is a solution of the Schr\"odinger equation
\begin{equation}
i \hbar \partial_t \hat U(t_1,t_0)=\hat H(t) \hat U(t_1,t_0),
\label{eq:2}
\end{equation}
with the initial condition $\hat U(t_0,t_0)=\hat 1$. Then, the conditional state of the system at time $t=t_1$ is given by
\begin{equation}\label{U}
\hat \rho (t_1,\mu)=\hat U(t_1,t_0) \hat \rho(t^+_0,\mu) \hat U^{\dagger}(t_1,t_0).
\end{equation}
At time $t=t_1$ it is performed a second projective measurement of the energy $\hat H(t_1)=\sum_{\nu } \nu | \nu \rangle \langle \nu |$, leading to the outcome $\nu$. Given the energy values obtained in the two projective measurements, one can compose the variable $W=\nu-\mu$. Due to the intrinsic randomness of the measurement process and the fact that the initial state is mixed, $W$ is a random variable. We denote $P_F(W)$ its corresponding probability distribution.

We now consider a second setting, the so called \emph{backward protocol}, in which at initial time $t=t_1$ the system has the Hamiltonian $\hat H(t_1)$ and it is prepared in a thermal state $\hat \rho_{th}(t_1)={\cal Z}(t_1)^{-1} \exp{(-\beta \hat H(t_1))}$. At this time value a first projective measurement of energy leads to the outcome $\nu'$. Then the system evolves in such a way that after a time interval $t_1-t_0$ the Hamiltonian becomes $\hat H(t_0)$, and a second energy measurement gives the result $\mu'$. The state of the system after this second measurement can be expressed as
\begin{align}
\hat \rho'(t_0)=P^{-1}(\mu',\nu')\hat \Pi_{\mu'} \hat U(t_0,t_1) \hat\Pi_{\nu'}  \hat \rho_{th}(t_1) \hat\Pi_{\nu'} \hat U^{\dagger}(t_0,t_1) \hat\Pi_{\mu'},
\label{eq:4}
\end{align}
with $P(\mu',\nu')$ the joint probability distribution of the outcomes $\mu'$ and $\nu'$. As in the forward protocol, the variable $W'=\mu'-\nu'$ is a random variable with probability distribution $P_B(W')$. 

The Fluctuation Theorem elucidates a subtle symmetry between the forward and backward probability distributions, $P_F$ and $P_B$, namely \cite{crooks1999a}
\begin{align}\label{eq:FTPM}
P_F(W)= e^{-\beta (F(t_1)-F(t_0)-W)} P_B(-W)=e^{-\beta(\Delta F -W)} P_B(-W)\,,
\end{align}
where $F=-k_B T \ln {\cal Z}$ is the Helmholtz free energy. From such relation it follows the Jarzynski equality \cite{jarzynski1997equilibrium,jarzynski1997nonequilibrium,jarzynski2011equalities} 
\begin{align}\label{eq:Jeq}
\Big \langle e^{\beta (\Delta F-W)} \Big \rangle=1\,.
\end{align}

In the above discussion, all energy measurements were projective so the state of the system just after each measurement was an eigenstate of the Hamiltonian. Therefore, according to the projection postulate, if two consecutive measurements were made, the resulting outcome energies would be the same. In this sense, measurements are reproducible. One may consider a more general setting in which the meter that monitors the system does not perform projective measurements but unsharp ones in which the outcomes are not reproducible, in the sense that two consecutive measurements will give different results \cite{diosi2011ashort}. 


A way to describe non-precise measurements is to consider the conditional probability $G(f|a)$ of certain outcome record $f$ if the system is in an eigenstate compatible with the eigenvalue $a$ \cite{braginsky2003quantum,wiseman2009quantum}. In general, such conditional probability is a function centered around $a$ and presents some width $\sigma$ that it is interpreted as the measurement error. Notably, there is a fluctuation theorem for those situations \cite{watanabe2014generalized} that incorporates the details of the measurement through the Fourier transform of the conditional probability  $G(f|0)$. In this context, Fluctuation Theorems for thermodynamics observables and dynamics described by complete positive trace preserving maps have been reported  \cite{albash2013fluctuation}.

In both the projective and non-projective measurement schemes, the work $W$ done on the system is defined as the difference of the outcomes of two energy measurements.

\section{Measurement and work}

Recently, definitions of work based on a single generalized energy measurement performed at an specific time value, and the subsequent unitary evolution of the post-measurement state have been proposed \cite{roncaglia2014work}. Such definitions are arguably more consistent from a thermodynamic point of view, especially if the cost of such measurement needs to be taken into account. Remarkably, the introduction of a more consistent definition of work during the measuring process leads to a generalized Jarzynski equality \cite{deffner2016quantum}, latter extended to open systems in \cite{sone2020quantum}.
In this work we consider non-projective measurements, and make explicit the role of the measuring apparatus and coherences in work fluctuations. 

To start with, we consider that the results of measurements when monitoring a system observable $\hat A=\sum_{a} a |a \rangle \langle a |$ can be represented by a continuous real variable $f$ which is located around the spectrum of $\hat A$ \cite{braginsky2003quantum,diosi1988cqm,braunstein1988quantum,breuer2002theory,wiseman2009quantum,sokolovski2019fromquantum}. Formally, such a variable can be defined in terms of a set of positive operators
\begin{equation}\label{Gmeasure}
\hat G^{1/2}(f|\hat A)\equiv \sum_{a} \sqrt{G(f|a)} |a \rangle \langle a |,
\end{equation} 
where the functions $G(f|a)$ are assumed to be real and non-negative in the variable $f$. We further assume that $G(f|a)=G(f-a)$. If a measurement of $\hat A$ on the state $\hat \rho(t)$ leads to the result $f$, the post-measurement state becomes
\begin{equation}
\label{poststate1}
\hat \rho(t^+,f)=P(f)^{-1} \hat G^{1/2}(f|\hat A)  \,\hat \rho(t)  \,\hat G^{\dagger 1/2}(f|\hat A)\,, 
\end{equation}
where
\begin{align}
P(f)=\text{Tr } \hat G^{1/2}(f|\hat A)  \,\hat \rho(t)  \,\hat G^{\dagger 1/2}(f|\hat A) \equiv \text{Tr } 
\hat G(f|A) \hat \rho(t)
\label{eq:Pf}
\end{align}
is the probability density function of obtaining the outcome $f$. In this sense, we consider unsharp measurements of the observable $\hat A$. The operators $\hat G^{1/2}(f|\hat A)$ satisfy $\int \, df \hat G^{\dagger 1/2}(f|\hat A) \hat G^{1/2}(f|\hat A)=\hat 1$, which ensures that $P(f)$ is normalized to one. 

We now consider that at a given time $t=t_0$ the system is in a state $\hat \rho(t_0)$. At such time value it is performed an energy measurement which leads to the outcome $f_0$. Then, our state of knowledge of the system changes abruptly and the post-measurement state $\hat \rho(t^+_0,f_0)$ becomes the normalized conditioned state given by Eq. (\ref{poststate1}).
After this energy measurement, the system undergoes a unitary evolution up to time $t=t_1$, dictated by the evolution operator $\hat U(t_1,t^+_0)$ corresponding to the Hamiltonian $\hat H(t)$. The final state of the system can be expressed as
\begin{equation}\label{Upoststate}
\hat \rho(t_1,f_0)=\hat U(t_1,t^+_0)\hat \rho(t^+_0,f_0)\hat U^{\dagger}(t_1,t^+_0).
\end{equation}

The outcome $f_0$ and the post-measurement conditioned state $\hat \rho(t_1,f_0)$ provide enough information to define work as the difference between the expectation value of energy at time $t=t_1$ and the measurement outcome $f_0$ \cite{deffner2010quantumwork}. Thus, the amount of work associated with the unitary evolution from $t_0^+$ to $t_1$ under the action of the external driving is defined as
\begin{equation}\label{work}
W(t_1,t_0^+,f_0)=\text{Tr }\hat H(t_1) \hat \rho(t_1,f_0)-f_0.
\end{equation}

\section{Probability density of work and Jarzynski equality}

The work $W(t_1,t^+_0,f)$ has a probability distribution $P_F(W)$ given by
\begin{equation}\label{FTformal}
P_F(W)=\int df \, P(f) \delta(W-W(t_1,t^+_0,f)),
\end{equation}
with $\delta(x)$ the Dirac delta distribution.

The average $\left\langle e^{-\beta W} \right\rangle$ is formally obtained from $P_F(W)$ as
\begin{equation}\label{averaexp}
\left\langle e^{-\beta W} \right\rangle =\int df \, P(f) e^{-\beta W(t_1,t^+_0,f)}.
\end{equation}
During the measurement process and the subsequent unitary evolution, the system changes its entropy and it generates coherences. Also, individual energy measurements exhibit fluctuations around the energy expectation value at the initial time. To make explicit these contributions to the fluctuation theorem, one can introduce a reference path of states $\rho_{th}(t)={ \exp{(-\beta H(t))}/{\cal{Z}}(t)}=\exp{(-\beta(H(t)-F(t))) }$, where $F(t)$ is the Helmholtz free energy. Then, working on equation (\ref{averaexp}), we reach
\begin{equation}
\left\langle e^{-\beta W} \right\rangle =\int df \, P(f) e^{\ln {\cal Z}(t_1)-S(\hat \rho(t_1,f)||\hat \rho_{th}(t_1))-S(\hat \rho(t_1,f))+\beta f},
\end{equation}
where $S(\hat X||\hat Y)=\text{Tr }\hat X \ln \hat X-\hat X \ln \hat Y$ is the quantum relative entropy between the states $\hat X$ and $\hat Y$ and $S(\hat X)=-\text{Tr } \hat X \ln \hat X$ is the von Neumann entropy of the state $\hat X$ \cite{nielsen2010quantum}. If one introduces $\Delta F=F(t_1)-F(t_0)$ as the difference between the Helmholtz free energies of the thermal states  $\hat \rho_{th}(t_1)$ and $\hat \rho_{th}(t_0)$, we arrive to the expression
\begin{equation}\label{FT1}
\left\langle e^{\beta(\Delta F-W)} \right\rangle= e^\xi
\end{equation}
with
\begin{equation}\label{FT2}
\xi=\ln \left\langle e^{\beta (f-\langle H(t_0) \rangle)} e^{-S(\hat \rho(t_1,f))+S(\hat \rho(t_0))} e^{-S(\hat \rho(t_1,f)||\hat \rho_{th}(t_1))+S(\hat \rho(t_0)||\hat \rho_{th}(t_0))} \right\rangle,
\end{equation}
where the averages are taken over $P(f)$.  Different effects such coherences, noise induced by measurements, information gained from measurement outcomes, all contribute to the Jarzynski equality (\ref{FT2}). To make their role explicit, we now analyze them separately.

\subsection{Coherences}

Coherence is one of the key elements of quantum physics and many of the most fundamental quantum phenomena \cite{streltsov2017quantum}. It is known that coherences play a role in measurement processes, considered as a change of information between two different basis \cite{groenewold1971aproblem}. Moreover, coherences are involved in the production of entropy in open quantum systems under non-equilibrium conditions as nicely discussed in \cite{santos2019therole}. 

Since the initial state of the system commutes with the initial Hamiltonian, the state immediately following an energy measurement will be diagonal in the eigenbasis of $\hat H(t_0)$. However, during the subsequent unitary evolution leading to state $\hat \rho(t_1,f)$, coherences will develop in the eigenbasis of the final Hamiltonian $\hat H(t_1)$. As we will show below, these coherences play a role in the Jarzynski relation (\ref{FT2}).

Given a Hamiltonian $\hat H(t)=\sum_\mu  |\mu(t)\rangle \mu(t) \langle\mu(t)|$, a dephasing operator relative to the basis $\{|\mu(t)\rangle\}$ and acting on a state $\hat \rho$ can be defined as
\begin{equation}
\Delta_{\hat H(t)} \hat \rho:=\hat \rho_D=\sum_\mu  |\mu(t)\rangle  \langle\mu(t)|\hat \rho |\mu(t)\rangle  \langle\mu(t)|.
\end{equation}
The action of this operator is to produce a state in the eigenbasis of $H(t)$ and erase all its non-diagonal elements. As a measure of quantum coherence we use the relative entropy of coherence $C_{H(t)}(\hat \rho)$ referred to $\hat H(t)$ \cite{baumgratz2014quantifying,streltsov2017quantum}, and given by  
\begin{equation}
C_{H(t)}(\hat \rho)=S(\hat \rho_D)-S(\hat \rho).
\end{equation}
In general coherence is a basis dependent concept. Here we select the basis that diagonalize the Hamiltonian at the given time and its associated thermal state. In a more extended setting, the role played by coherences in non-equilibrium entropy production has been analyzed in detail in \cite{santos2019therole}.

It can be seen that the quantum relative entropy $S(\hat \rho(t,f)||\hat \rho_{th}(t))$ can be decomposed into two contributions, one due to coherences and another associated with the populations. This last contribution measures the observer's ability to realize that after many measurements the populations of the final state are given by the state $\hat \rho_D(t,f)$, and not by the thermal state $\hat \rho_{th}(t)$ \cite{witten2018anintroduction}. The coherences are measured by the relative entropy of coherence $C_{H(t)}(\hat \rho)$, whereas the contribution due to the populations is expressed in terms of the Kullback-Leibler divergence \cite{nielsen2010quantum}
\begin{equation}\label{dkl}
D_{KL}(\hat \rho_D(t,f)||\hat \rho_{th}(t)):=D_{KL}(\{p_{\mu}\}||\{q_{\mu}\})=\sum_{\mu} p_{\mu} \ln \frac{p_{\mu}}{q_{\mu}}\,,
\end{equation}
with $p_{\mu}= \langle\mu(t)|\hat \rho_D(t,f) |\mu(t)\rangle$ and $q_{\mu}= \langle\mu(t)|\hat \rho_{th}(t) |\mu(t)\rangle$ the probability distributions associated with populations of the states $\hat\rho_D(t,f)$ and $\hat\rho_{th}(t)$, respectively. Notice that the states $\hat \rho_D(t,f)$ and $\hat \rho_{th}(t)$ are diagonal in the eigenbasis of $\hat H(t)$ so they refer to populations in such basis.

Thus, the quantum relative entropy can be written as
\begin{equation}\label{REntropy}
	S(\hat \rho(t,f)||\hat \rho_{th}(t))=C_{H(t)}(\hat \rho(t,f))+D_{KL}(\hat \rho_D(t,f)||\hat \rho_{th}(t))\,,
\end{equation}
which makes explicit the role of coherences in the Jarzynski equality (\ref{FT1}). In particular, at the initial time
\begin{equation}\label{RelativeEntropyini}
S(\hat \rho(t_0)||\hat \rho_{th}(t_0))=C_{H(t_0)}(\hat \rho(t_0))+D_{KL}(\hat \rho_D(t_0)||\hat \rho_{th}(t_0))\,,
\end{equation}
which becomes zero in the case of an initial thermal state $\hat \rho(t_0)=\hat \rho_{th}(t_0)$.


\subsection{Information gained by the observer in a measurement}

%

The term corresponding to the difference of von Neumann entropies $\Delta S=S(\hat \rho(t,f))-S(\hat \rho(t_0))$ is related to the amount of information involved in the measurement process. We start by considering the average of $\Delta S$ over the probability distribution $P(f)$, such that
\begin{equation}\label{Incentropy}
P(f)\Delta S=P(f)\ln P(f)-\text{Tr }\hat G(f|\hat H(t_0))\hat \rho(t_0)\ln \hat G(f|\hat H(t_0))-\text{Tr }\Big(\hat G(f|\hat H(t_0))-P(f)\Big)\hat\rho(t_0)\ln\hat \rho(t_0).
\end{equation}
For simplicity we have assumed that the initial state commutes with $H(t_0)$.

Then, we introduce the Shannon entropy of a distribution $g(f)$ \cite{nielsen2010quantum}
\begin{equation}
H_S(g(f))=-\int df \, g(f) \ln g(f),
\end{equation}
and integrate Eq. (\ref{Incentropy}) over all possible outcomes $f$. The resulting average entropy increase is given by 
\begin{equation}\label{shannon}
\langle \Delta S \rangle=-H_S(P(f))+H_S(G(f|0)),
\end{equation}
with $0\le \langle \Delta S \rangle \le S(\rho(t_0)) $.



It is natural that the quantity $\langle \Delta S \rangle$ is bounded by the entropy of the initial state when the observer collects its measurements. In the case that the post-measurement state is a pure state, such that $\langle \Delta S \rangle=S(\rho(t_0))$, the observer can infer the initial state from a sufficient large set of measurements. On the other hand, if $\langle \Delta S \rangle \le S(\rho(t_0))$ the observer may get outcomes that do not add information about the state after the measurement.



Thus, the term $\Delta S$ is related to the difference of two Shannon entropies for the probability distributions $P(f)$ and $G(f|0)$, which suggest its relation to the net amount of information involved in the measuring process.
A more detailed analysis, taking into account the accuracy of the meter, can be done to give meaning to Eq. (\ref{shannon}) in terms of information \cite{gaspard2013entropy}.

Notice that this last term is relevant in the case of non-projective measurements, and it contains the fact that some measurements can be more informative that others.

\subsection{Jarzynski equality}

Once we have analyzed the different terms involved in the Jarzynski identity (\ref{FT2}), it can be expressed in a more appealing form as
\begin{equation}\label{FT3}
\xi=\ln \left\langle e^{\beta ( f-\langle H(t_0) \rangle)} e^{-\Delta S} e^{-\Delta C}e^{-\Delta D_{KL}} \right\rangle,
\end{equation}
where
\begin{equation}\label{CDKL}
\Delta C=C_{H(t_1)}(\hat \rho(t_1,f))-C_{H(t_0)}(\hat \rho(t_0))
\end{equation}
gives the change in the amount of coherences during the measurement process with respect to coherences of the known initial state, and
\begin{equation}\label{DKL}
\Delta D_{KL}=D_{KL}(\hat \rho_D(t_1,f)||\hat \rho_{th}(t_1))-D_{KL}(\hat \rho_D(t_0)||\hat \rho_{th}(t_0)) ,
\end{equation}
is the difference of the final and initial Kullback-Leibler divergences. As an initial hypothesis, the thermal state $\hat \rho_{th}(t)$ can be used by the observer as a basis for predicting the probability of observing a given measurement result. However, it may well be that a different state $\hat \rho$ leads to better predictions. 
The Kullback-Leibler divergence measures whether such initial hypothesis is wrong. Specifically, if after a sufficiently large number $n$ of measurements it follows that $n D_{KL}(\hat \rho||\hat \rho_{th}) \ll 1 $, then the observer can conclude that the state $\hat \rho_{th}(t)$ is not a good representation of the observed statistics. Thus,  Eq. (\ref{DKL}) measures how wrong the observer will be if the outcome statistics are assumed to be given by thermal states.

Finally, the Jarzynski identity (\ref{FT2}) establishes a bound on the average work done on the system. Taking into account the Jensen inequality \cite{jensen1906surlesfonctions} it follows   
\begin{align}\label{SL}
\beta \left \langle W \right \rangle \ge \beta \Delta F-\xi\,,
\end{align}
which can be considered as a generalization of the second law \cite{jarzynski2011equalities}. This last expression provides a generalization of an analogous relation obtained for projective measurements \cite{deffner2016quantum} to the case of non-orthogonal measurements, insofar as $\xi$ includes information regarding the resolution of the apparatus that monitors the energy of the system. We analyze this issue in more detail in the illustrative example presented below. 



\section{Two level system}


We now consider a non-trivial example to illustrate the essential elements discussed in the previous sections. Specifically, we consider a system of two energy levels, which is described by the time-dependent Hamiltonian 
\begin{equation}\label{hamiltoniantl}
\hat H(t)=\sum_{i=1,2}  |\mu_i(t)\rangle \mu_i(t) \langle\mu_i(t)|\,,
\end{equation} 
in terms of which we can write
\begin{align}
\hat G^{1/2}(f|\hat H(t_0))=\sum_{i=1,2}\sqrt{G(f|\mu_i(t_0))} |\mu_i(t_0)\rangle \langle\mu_i(t_0)|.
\end{align}
We remind that the functions $G(f|\mu_i(t_0))=G(f-\mu_i(t_0))$. We also assume that such functions exhibit a characteristic width of order $\sigma$ around $\mu_i(t_0)$, and tend to zero as $|f-\mu_i(t_0)|$ becomes large enough. Then, the probability density function Eq. (\ref{eq:Pf}) for an initial state that commutes with the initial Hamiltonian, becomes
\begin{align}
P(f)=\sum_{i=1,2} G(f-\mu_i(t_0)) \langle \mu_i(t_0)| \hat \rho(t_0) |\mu_i(t_0)\rangle \equiv \sum_{i=1,2} G(f-\mu_i(t_0)) p_i\,.
\end{align}
The conditional state Eq. (\ref{poststate1}), just after performing an energy measurement with outcome $f$ at time $t_0$, is given by
\begin{align}
\hat\rho(t_0^+|f)=P^{-1}(f) \sum_{i=1,2} p_i G\left(f-\mu_i(t_0)\right)  |\mu_i(t_0)\rangle \langle\mu_i(t_0)|\,.
\end{align}
Whereas, according to Eq. (\ref{Upoststate}), the state after the unitary evolution up to the final time $t_1$ can be written as 
\begin{align}\label{poststate2}
\hat \rho(t_1|f)&=\sum_{i=1,2} p_i \frac{G\left(f-\mu_i(t_0)\right)}{P(f)} \hat U(t_1,t^+_0)|\mu_i(t_0)\rangle \langle\mu_i(t_0)|\hat U^{\dagger}(t_1,t^+_0) \nonumber \\
&\equiv \sum_{i=1,2} p_i \frac{G\left(f-\mu_i(t_0)\right)}{P(f)} \hat \rho_i(t_1) 
\equiv \sum_{i=1,2} p_i \tilde G\left(f-\mu_i(t_0)\right) \hat \rho_i(t_1).
\end{align}

\begin{figure}[h]\centering
\includegraphics[width=0.5\linewidth]{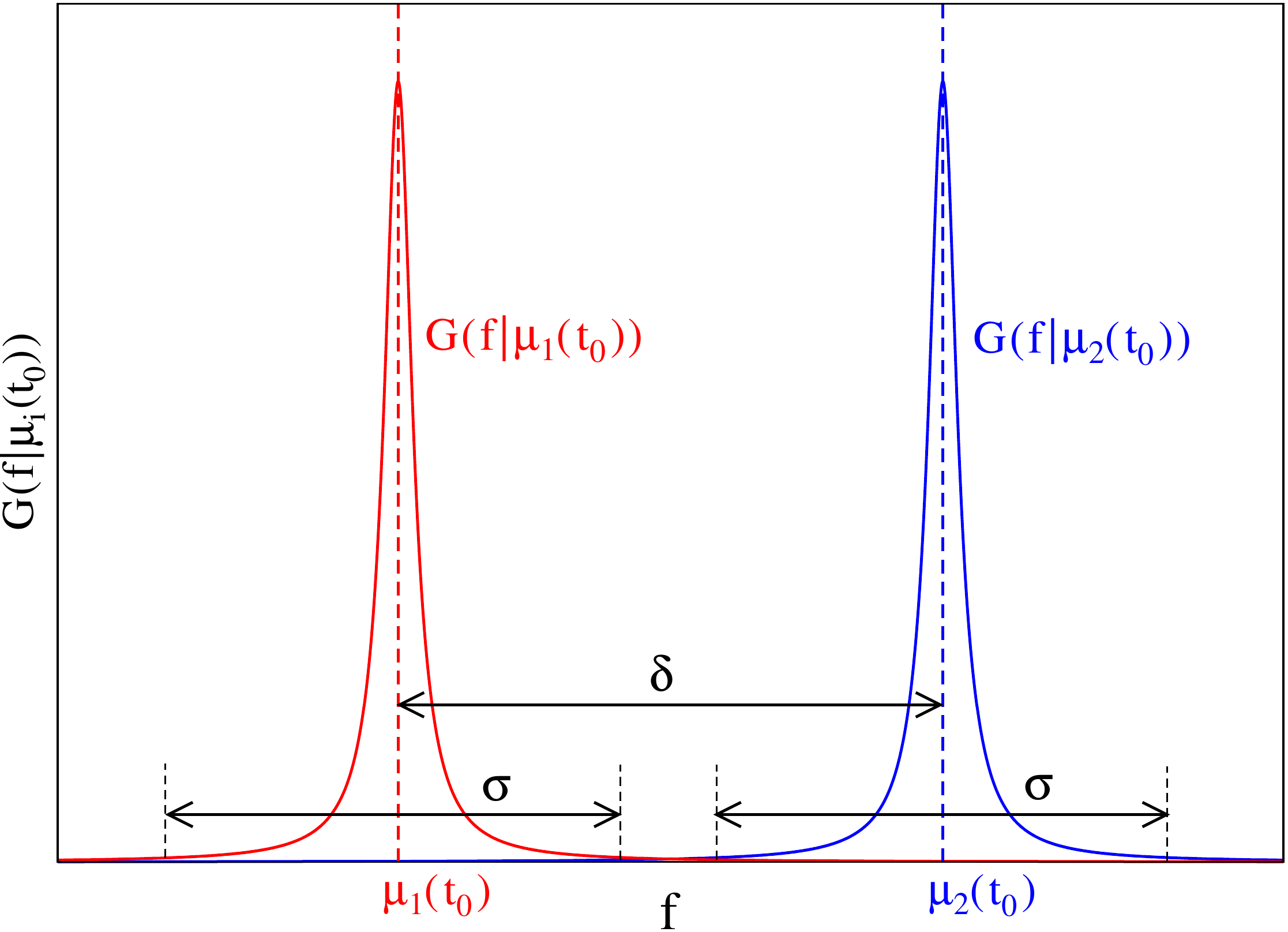}
\caption{A scheme of a possible distribution of $G\left(f|\mu_1(t_0)\right)$ and $G\left(f|\mu_2(t_0)\right)$, as functions of the outcome of the measurement $f$.}
\label{fig1}
\end{figure}

Two different scenarios arise depending on whether the two functions $G\left(f-\mu_1(t_0)\right)$ and  $G\left(f-\mu_2(t_0)\right)$ overlap or not, see Fig. \ref{fig1}.
On one hand, if $|\mu_1(t_0)-\mu_2(t_0)| \gg \sigma$ the functions $G\left(f-\mu_1(t_0)\right)$ and $G\left(f-\mu_2(t_0)\right)$ do not overlap. In this case, if the result of an energy measurement at time $t_0$ is $f$, the observer finds out that immediately after such measurement the system is in the state $|\mu_1(t_0)\rangle \langle\mu_1(t_0)|$ if $f$ is located within the domain of $G\left(f-\mu_1(t_0)\right)$ or in the state $|\mu_2(t_0)\rangle \langle\mu_2(t_0)|$ otherwise. In this sense we say that the measurement is informative since the outcome $f$ reveals which of the two possible energy states the system is in.  On the other hand, if $f$ is within the domain where there is significant overlap of both $G$-functions, then the post-measurement state will be a linear combination of the states $\lbrace |\mu_i(t_0)\rangle \langle\mu_i(t_0)| \rbrace\,(i=1,2)$. In this case we refer to non-informative measurements. The two possible scenarios are analyzed in detail below.


\subsection{Non-overlapping case}


In this case all measurements are informative, and the state just after the measurement will be either $|\mu_1(t_0)\rangle \langle\mu_1(t_0)|$ or $|\mu_2(t_0)\rangle \langle\mu_2(t_0)|$. If $J_i$ is the domain in which $G\left(f-\mu_i(t_0)\right)$ is appreciably different from zero, and $J_1 \cap J_2 =\emptyset$, then
\begin{align}
\left\langle e^{-\beta W} \right\rangle =\sum_{i=1,2} \int_{J_i} df \, p_i  G\left(f-\mu_i(t_0)\right) e^{-\beta {\text Tr } \hat H(t_1) \hat \rho_i(t_1)+ \beta f}\,.
\end{align}
Notice that the extension of the above expression to a system with a larger number of energy levels is straightforward. It can be rewritten as
\begin{align}\label{non_over}
e^\xi=\bar G(i \beta) \sum_{i=1,2}\left(p_i {\cal{Z}}(0) 
e^{\beta \mu_i(t_0)}\right)
e^{-S(\hat \rho_{ i}(t_1)||\hat \rho_{\text{th}}(t_1))}\,,
\end{align}
where we have introduced the characteristics of the measurement procedure, which are given by the Fourier Transforms of the $G(f)$ functions, $i.e.$ $\bar G(u)=\int df \, e^{-i u f} G(f)$. Equations (\ref{FT1}) and (\ref{non_over}) extend the fluctuation theorem derived in \cite{deffner2016quantum} to the generalized measurements discussed in previous sections and initial states commuting with the initial Hamiltonian. In the particular case of thermal initial states, the above expression reduces to
\begin{align}\label{jarzynskynonoverlap}
e^\xi=\bar G(i \beta) \sum_{i=1,2} e^{-S(\hat \rho_{i}(t_1)||\hat \rho_{\text{th}}(t_1))}\,.
\end{align}
The fluctuation relation obtained in this case is the one that would be obtained in projective measurements, multiplied by the function $\bar G(i \beta)$, which characterizes the noise induced by non-projective measurements. In a two-point measurement protocol such factor takes the form $|\bar G(i \beta) |^2$ due to the two measurements performed \cite{watanabe2014generalized}. Here, coherences play no role in the resulting Jarzynski equality.



\subsection{Overlapping case}


A more subtle situation arises in the case that $|\mu_1(t_0)-\mu_2(t_0)|\ll\sigma$, as there may be measurements in which the post-measurement state is a  linear combination of  $|\mu_1(t_0)\rangle \langle\mu_1(t_0)|$ and $|\mu_2(t_0)\rangle \langle\mu_2(t_0)|$. Here we denote by $I_1$ the domain in which the function $G(f-\mu_1{(t_0)})$ has values different from zero and the function $G(f-\mu_2{(t_0)})$ is zero or almost zero. The domain $I_2$ is defined in a similar way. Thus, in such domains there is no appreciable overlap between the two $G$-functions. We call the domain in which such overlap takes place $I_{ov}$. Then, we can write
\begin{equation}\label{overlap1}
\left\langle e^{-\beta W} \right\rangle =\Big(\int_{I_1} df+\int_{I_2} df+\int_{I_{ov}} df\Big) \, P(f) e^{-\beta {\text{Tr } \hat H(t_1) \hat \rho(t_1,f) }+\beta f}\,.
\end{equation}
If the outcome $f$ of the measurement is located within a non-overlapping domain $I_i$, with $i=1$ or $2$, the resultant state will be $|\mu_i(t_0)\rangle \langle\mu_i(t_0)|$, and the corresponding value of the probability density becomes $P(f)=p_i G(f-\mu_i(t_0))$. Whereas, a value of $f$ located within the overlapping region $I_{ov}$ leads to a post-measurement state given by Eq. (\ref{poststate2}). Thus, Eq. (\ref{overlap1}) can be rewritten as
\begin{equation}
\left\langle e^{-\beta W} \right\rangle=\sum_{i=1,2}\int_{I_i} df \, p_i\,G(f-\mu_i(t_0)) e^{-\beta {\text{Tr } \hat H(t_1) \hat \rho_i(t_1) }+\beta f}+
\int_{I_{ov}} df\,P(f) e^{-\beta {\text{Tr } \hat H(t_1) \hat \rho(t_1,f) }+\beta f}.
\end{equation}
Considering that 
\begin{equation}
\int_{I_i \cup I_{ov}} df \, p_i\,G(f-\mu_i(t_0)) e^{-\beta {\text{Tr } \hat H(t_1) \hat \rho_i(t_1) }+\beta f} = \int_{J_i}  df \, p_i\,G(f-\mu_i(t_0)) e^{-\beta {\text{Tr } \hat H(t_1) \hat \rho_i(t_1) }+\beta f}\,,
\end{equation}
it follows
\begin{align}\label{eq:ov1}
\left\langle e^{-\beta W} \right\rangle 
&=\sum_{i=1,2} \int_{J_i} df \, p_i\,G(f-\mu_i(t_0)) e^{-\beta   {\text{Tr }} \hat H(t_1)  \hat \rho_i(t_1) +\beta f} \nonumber \\
&+\int_{I_{ov}} df \, \Big( P(f) e^{-\beta \sum_{i} p_i \tilde G\left(f-\mu_i(t_0)\right) \text{Tr } \hat H(t_1)\hat \rho_i(t_1)+\beta f }-\sum_{i=1,2} p_i\,G(f-\mu_i(t_0)) e^{-\beta {\text{Tr } \hat H(t_1) \hat \rho_i(t_1) +\beta f}} \Big),
\end{align}
where we have used Eq. (\ref{poststate2}). The first element in this expression leads to the fluctuation expression obtained in the non-overlapping case. The second part gives the contribution due to the fact that measurements may be non-informative. Introducing the function
\begin{align}
\bar G_i(i \beta)=\int_{I_{ov}} df \, G(f-\mu_i(t_0)) e^{\beta (f-\mu_i(t_0))}
\end{align}
into Eq. (\ref{eq:ov1}), it follows that
\begin{align}\label{eq:ov2}
	e^{\xi}&=\sum_{i=1,2} \left(p_i {\cal{Z}}(0) e^{\beta \mu_i(t_0)}\right) \Big(\bar G(i \beta)-\bar G_i(i \beta)\Big) 
	e^{-S(\hat \rho_i(t_1)||\hat \rho_{\text{th}}(t_1))} \nonumber \\
	&+ \int_{I_{ov}} df \, \Big( \sum_{i=1,2} p_i {\cal{Z}}(0) G\left(f-\mu_i(t_0)\right)  e^{\beta f}\Big)
	e^{-\sum_i p_i \tilde G(f-\mu_i(t_0))S(\hat \rho_i(t_1)||\hat \rho_{\text{th}}(t_1))}\,.
\end{align}
We remind that $\tilde G(f-\mu_i(t_0))= G\left(f-\mu_i(t_0)\right)/P(f)$. As mentioned above, $p_i {\cal{Z}}(0) e^{\beta \mu_i(t_0)}=1$ for an initial thermal sate.

The last expression establishes the fluctuation relation for energy measurements in which it is not possible to conclude that the post-measurement state is an eigenstate of the Hamiltonian being monitored. In contrast to the non-overlapping case, the fluctuation relation can now not be expressed as the product of a term corresponding to projective measurements and one term involving the characteristics of the measurement procedure. In this sense, the information concerning the system and the noise induced by measurements appear mixed in the fluctuation relation for the overlapping case.  For further discussion of this issue, details concerning the measurement procedure, such as the $G(f)$ function, should be analyzed in order to provide a more explicit expression for the last term in Eq. (\ref{eq:ov2}).

\section{Illustrative example}

In general, the conditional probability $G(f|a)$ is a function with appreciable non zero values in the variable $f$ within a domain of size of order $\sigma$ around the value $a$. In the case that $G$ is a Gaussian function, $\sigma$ would correspond to several times its width.  



To illustrate the analysis shown in the previous sections we consider the following form of the function $G$
\begin{equation}\label{heaviside}
G(f|a)=
\begin{cases}
\sigma^{-1} & f \in (a-\sigma/2,a+\sigma/2) \\
0 &  \text{ otherwise}, 
\end{cases}
\end{equation}
and a two level system with initial and final Hamiltonians $\hat H(t_0)$ and $\hat H(t)$ respectively, see Eq. (\ref{hamiltoniantl}). The initial state is taken to be $\hat \rho_{th}(t_0)={{\cal{Z}}^{-1}(t_0)} \sum_{i=1,2} e^{-\beta \mu_i(t_0)} |\mu_i(t_0) \rangle \langle \mu_i(t_0)|$. The probability density $P(f)$ of measurement outcome $f$ is given by
\begin{equation}\label{Pf}
P(f)={\cal{Z}}^{-1}(t_0) \left( e^{-\beta \mu_1(t_0)} G(f|\mu_1(t_0))+e^{-\beta \mu_2(t_0)} G(f|\mu_2(t_0)) \right)=p_1 G(f|\mu_1(t_0))+ p_2 G(f|\mu_2(t_0)).
\end{equation}

As mentioned above, two interesting cases arise. In the case of non-overlapping, in which $\sigma < \left|\mu_2(t_0)-\mu_1(t_0)\right|$, a measurement corresponding to a given outcome $f$ updates the observer state of knowledge. Here the normalized post-measurement state will be one of the eigenstates of the observable measured $\hat H(0)$. Whereas if overlap occurs, with $\sigma > |\mu_2(t_0)-\mu_1(t_0)|$, those measurement outcomes that lie in the region of overlap between $G(f|\mu_1(t_0))$ and $G(f|\mu_2(t_0))$ do not provide a gain of information. In this case the state of knowledge of the observer will not be updated and neither will be the quantum state. We analyze both cases in more detail below.

\subsection{Non-overlapping case}

The post-measurement state is given by a two-piecewise density matrix, defined within the intervals $I_1=\{ f |\mu_1(t_0)-\sigma/2<f<\mu_1(t_0)+\sigma/2\}$ and $I_2=\{ f |\mu_2(t_0)-\sigma/2<f<\mu_2(t_0)+\sigma/2\}$ $i.e.$
\begin{equation}
\hat \rho(t_0^+,I_1)= |\mu_1(t_0) \rangle \langle \mu_1(t_0)|  \quad \text{and }\quad \hat \rho(t_0^+,I_2)= |\mu_2(t_0) \rangle \langle \mu_2(t_0)|.	
\end{equation}
The probability of an outcome $f$ is given by Eq. (\ref{Pf}), and the state at time $t_1$ is given by
\begin{equation}
\hat \rho(t_1,I_i)=\hat U(t_1,t_0^+) \hat \rho(t_0^+,I_i)\hat U^{\dagger}(t_1,t_0^+), \quad i\in\{1,2\}.
\end{equation}
From the state of the system at final time, the probability distribution $P(f)$ and Eq. (\ref{FT2}) it follows that
\begin{align}\label{Jnonov}
	e^{\xi}=\frac{2}{\beta \sigma} \sinh (\beta \sigma /2)\sum_{i=1,2} e^{-S(\hat \rho(t_1,I_i)||\hat \rho_{th}(t_1))}.
\end{align}


The first factor shows the influence of the measuring device on the work fluctuations. Considering $\bar G(u)=\int df \, e^{-i u f} G(f|0)$, such factor is just $\bar G(i \beta)$. 
The second factor corresponds to ideal von Neumann measurements  \cite{deffner2016quantum}.

So we can conclude that the fluctuation theorem for work, as defined in Eq. (\ref{work}), when extended to non-projective measurements, see Eq. (\ref{Gmeasure}), is modified by a prefactor containing the effect of measurements.





\subsection{Overlapping case}

A more intriguing situation arises in the overlapping case. Here there is an overlapping region for possible measurement outcomes, $I_{ov}=\{f|\mu_2(t_0)-\sigma/2<f<\mu_1(t_0)+\sigma/2\}$, in which no information is gained by the observer. If $I_1=\{ f |\mu_1(t_0)-\sigma/2<f<\mu_2(t_0)-\sigma/2\}$ and $I_2=\{ f |\mu_1(t_0)+\sigma/2<f<\mu_2(t_0)+\sigma/2\}$, the corresponding normalized post-measurement states with outcomes within each of those regions are given by 
\begin{equation}
\hat \rho(t_0^+,I_1)=|\mu_1(t_0) \rangle \langle \mu_1(t_0)|\,,\,\,\, \hat \rho(t_0^+,I_2)= |\mu_2(t_0) \rangle \langle \mu_2(t_0)|\,\,\mbox{and}\,\,\, \hat \rho(t_0^+,I_{ov})=\hat \rho_{th}(t_0),
\end{equation}
respectively. The state after the unitary evolution up to time $t_1$ takes the form
\begin{equation}
\hat \rho(t_1,I_i)=\hat U(t_1,t_0^+) \hat \rho(t_0^+,I_i)\hat U^{\dagger}(t_1,t_0^+)\,,
\end{equation} 
with $i\in\{1,2,ov\}$.
 
In this case it follows that
\begin{equation}
	S(\hat \rho(t_1,I_{ov})||\hat \rho_{th}(t))=-S(\rho(t_1,I_{ov}))+p_1 S_{re}(1)+p_2 S_{re}(2)\,.
\end{equation}
To shorten the expressions we have introduced the following notation, $S_{re}(i)=S(\hat \rho_i(t_1)||\hat \rho_{th}(t))$ and $p_i={{\cal{Z}}^{-1}(t_0)}e^{-\beta \mu_i(t_0)}$, for $i\in \{1,2\}$.  

Here, following the same steps as in the previous section, it can be concluded that
\begin{align}\label{FT4}
	e^{\xi}&=\frac{2}{\beta \sigma} \sinh (\beta \sigma /2)\sum_{i=1,2}e^{-S_{re}(i)} \nonumber \\	&+\frac{2}{\beta \sigma} \Big(\sinh \left(\beta \left(\mu_2(t_0)-\mu_1(t_0)\right)-\beta\sigma/2\right)-\sinh (\beta \sigma /2)\Big) \Big(p_1 e^{-S_{re}(1)} +p_2 e^{-S_{re}(2)}-e^{-p_1 S_{re}(1)-p_2 S_{re}(2)}\Big).
\end{align}

In this expression one can identify the contribution due to the non-overlapping case, and the correction that arises due to uninformative measurement records. In contrast to what happens in the non-overlapping case, it is striking that here the effect of the measuring device is not a global factor that multiplies a term that depends only on the system. In this situation, in which the observer is confronted with measurement outcomes that do not contribute to update her/his state of knowledge about the system, Jarzynski's equality exhibits a mixture of information concerning the system and that corresponding to the measuring apparatus. The last term in Eq (\ref{FT4}) contains the fluctuations of the relative entropies in the regions $I_1$ and $I_2$ and is a correction to fluctuation relation (\ref{jarzynskynonoverlap}). Indeed, the functions $S_{re}(i)$ are constant within the intervals $I_i$ so we can write
\begin{equation}\label{flucs1s2}
p_1 e^{-S_{re}(1)} +p_2 e^{-S_{re}(2)}-e^{-p_1 S_{re}(1)-p_2 S_{re}(2)}:= \Big< e^{-S_{re}} \Big>- e^{- \langle S_{re} \rangle}.
\end{equation}
It is worth analyzing the measurement regime, in which $\sigma \gg |\mu_2(t_0)-\mu_1(t_0)|$, the expression for $e^{\xi}$ factorizes in a term that depends only on the meter characteristics and the temperature, and another term that does not contain any information about the meter. In such regime
\begin{align}
		e^{\xi}\approx \frac{2}{\beta \sigma} \sinh (\beta \sigma /2) \Big(\sum_{i=1,2}e^{-S_{re}(i)}+ 2(\Big< e^{-S_{re}} \Big>- e^{- \langle S_{re} \rangle}) \Big).
\end{align}

On the other hand, in the low temperature regime the probabilities $p_1 \approx 1\,(p_2\approx 0)$, then the quantity (\ref{flucs1s2}) approaches to zero and expression (\ref{Jnonov}) is recovered.

This result is an illustration of Eq.(\ref{FT2}) as incorporates the contribution of non-informative measurements to Jarzynski equality derived in \cite{deffner2016quantum}.

\subsection{Numerical simulations}

To numerically simulate the measurement process we have considered the characteristic function (\ref{heaviside}) and the dynamics of a two level system described by the Hamiltonian
\begin{align}\label{HTL}
H(t)=\hbar \omega_q \frac{\sigma_z}{2}+\hbar \Omega_R \sigma_y \cos \left(\omega_q t+\psi\right),
\end{align}
with $\sigma_\alpha$ ($\alpha=y,z$) the Pauli matrices, $\omega_q$ the resonant frequency, $\psi$ a phase and $\Omega_R$ the Rabi frequency associated with the driving. This system has been implemented in a nice experiment \cite{naghiloo2020heat} to study individual quantum state trajectories and its properties. Specifically, it has been shown that it is possible to study quantum thermodynamic properties for individual quantum trajectories, and that the results obtained from the master equation approach are consistent with the two-projective-measurement scheme. 

\begin{figure}[h]\centering
\includegraphics[width=0.7\linewidth]{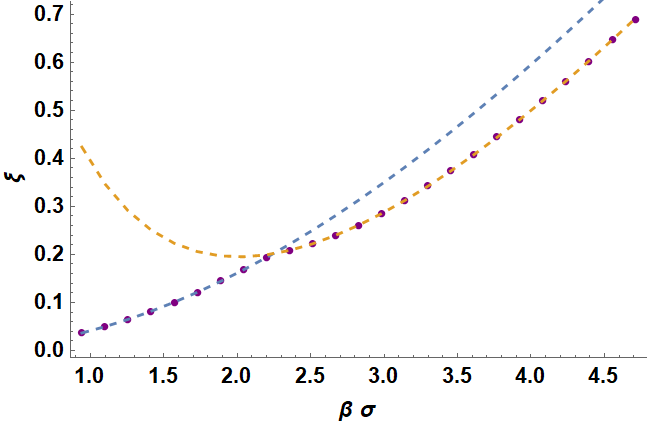}
\caption{
The evolution of $\xi$ {\sl vs} $\beta\sigma$ in a measurement process with characteristic function (\ref{heaviside}), in a two level system described by the Hamiltonian (\ref{HTL}). The dots are the result of the numerical simulation. The blue dashed line corresponds to the theoretical expressions, (\ref{jarzynskynonoverlap}) and (\ref{Jnonov}), for the non-overlapping case, and the orange dashed line to the expressions, (\ref{eq:ov2}) and (\ref{FT4}), for the overlapping case. The system parameters, $\omega_q=2 \pi \times 6.541 \times 10^9 \text{Hz}$ and $\Omega_R= 2 \pi \times 10^6 \text{Hz}$, have been taken from \cite{naghiloo2020heat}. The temperature is $T=0.14 \text{K}$, the final time $t_f=2 \pi/3 \omega_q$ and $\psi=\pi/4$. }
\label{figure1}
\end{figure}

In the numerical simulations we have considered as initial state the one corresponding to the canonical thermal state of the Hamiltonian $H(0)$, and a measurement process with characteristic function (\ref{heaviside}). Figure \ref{figure1} shows the agreement between the numerical results and those obtained from the analytical expressions derived in the previous sections. In particular, the transition between the non-overlapping and overlapping cases can be clearly observed. Note that this transition would be smoother if $G(f|0)$ were a smooth function, such as a Gaussian. In Figure \ref{figure1}, the difference between the blue and orange lines evidences the contribution of the region of overlap, given by the second term of equation (\ref{FT4}).

\begin{figure}[h]\centering
\includegraphics[width=0.7\linewidth]{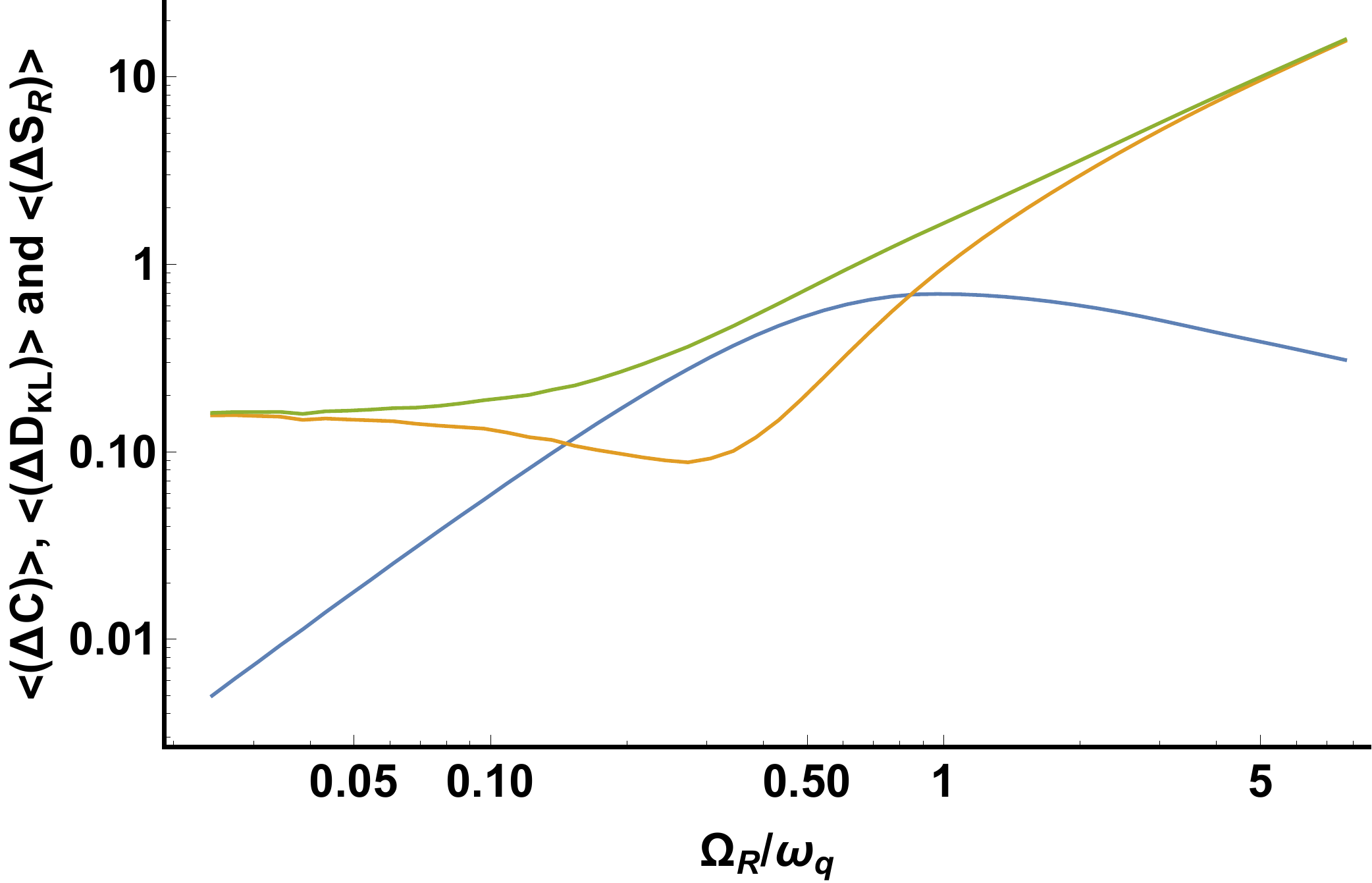}
\caption{Numerical results for the mean relative entropy of coherence $\langle\Delta C\rangle=\langle C_{H(t)}(\hat\rho(t,f))-C_{H(t_0)}(\hat\rho(t_0))\rangle$ (blue line), the mean increment of Kullback-Leibler divergence $\langle\Delta D_{KL}\rangle=\langle D_{KL}(\hat \rho_D(t,f)||\hat\rho_{th}(t))-D_{KL}(\hat \rho_D(t_0)||\hat\rho_{th}(t_0))\rangle$ (orange line), and mean increment of the relative entropy $\langle\Delta S_R\rangle=\langle S(\hat\rho(t,f)||\hat\rho_{th}(t_0))-S(\hat\rho(t_0)||\hat\rho_{th}(t_0))\rangle$ (green line) vs. the frequency ratio  $\Omega_R/\omega_q$. Notice how the increment of the relative entropy of coherence saturates at a value $\ln 2$ (see text).  The resonant frequency has been set at $\omega_q=2 \pi \times 6.541 \times 10^9 \text{Hz}$. The temperature is $T=0.14 \text{K}$, the final time $t_f=2 \pi/3 \omega_q$, and $\psi=0$ and $\sigma=2 \hbar \omega_q$. The averages $\langle*\rangle$ are taken over $10^5$ realizations.}
\label{fig2}
\end{figure}

Due to the initial state chosen in the measurement process, the initial relative entropy, the relative entropy of coherence and the Kullback-Leibler divergence are all zero. However, once the measurement occurs and the subsequent time evolution takes place the system will develop coherences due to the external driving, and will reach the final time with non-zero Kullback-Leibler divergence.

The relative entropy of coherence is expected to grow as the Rabi frequency increases. However, for large values of $\beta \Omega_R/\omega_q$ the coherences decrease until they decay to zero. In this limit the initial and final thermal states, the post-measurement state and the evolution operator all commute and no coherences develop. At the strict limit of no driving, in which $\Omega_R/\omega_q \to 0$, the increment of the entropy equals the von Neumann entropy of the initial state, see Fig. \ref{fig2}. Additionally, the range of $\sigma$  values analyzed is such that small values of $\Omega_R/\omega_q$ correspond to the overlapping case. Therefore, the relative entropy is given by $(p_1 S_{re}(1)+p_2 S_{re}(2))/\sigma $. While in the case of large values of $\beta \Omega_R/\omega_q$ the mean increment of the relative entropy behaves as $\Delta S_R \approx \Delta D_{KL}\approx 2 \beta \hbar \Omega_R \cos(\omega_q t_f+\psi)$.

\section{Conclusions}

In this work we have analyzed the Jarzynski equality based on a protocol of a single energy measurement proposed in \cite{deffner2016quantum} for non-projective measurements. In this scheme the coherences developed in the system during the post-measurement time evolution play an important role. A comparative study of the resolution of the measurement device with respect to the distance between neighboring energy eigenvalues of the system allows different scenarios to be introduced.

If the energy resolution of the meter can resolve the individual elements of the energy spectrum, then the resulting Jarzynski equality is that which would be obtained in projective measurements except for a multiplicative factor that depends on the meter. In this case the resulting signal can be convoluted to extract the one corresponding to the Jarzynski equality. On the other hand, if the meter cannot resolve a part of the energy spectrum, then the effect of the measurement is not a factor that multiplies the Jarzynski relation. 

We have analyzed a simple model of a two-level system in which, although informative measurements appear due to the bad energy resolution of the meter, in the weak measurement limit it is possible to extract a new Jarzynski equality containing an additional term related to the non-projective measurements used.

We have also studied the behavior of coherences in a driven two-level system by means of the relative entropy coherence and the relative entropy change during a single measurement protocol. Numerical results show that the coherences grow with the Rabi frequency until they reach a maximum, and then decay to zero as this frequency becomes larger. Also, the change in the relative entropy of the post-measurement state at final time and the reference thermal state increases linearly with the Rabi frequency. 

To conclude, we have shown that there are realistic systems in which it is possible to observe the corrections to the Jarzynski equality based on a protocol of a single energy measurement.

\section{Acknowledgements}
We thank L. Cereli, L. Correa, G. De Chiara, M. Lostaglio, G. Manzano, J.G. Muga and M. Paternostro for fruitful discussions. Financial support of MINECO and the European Regional Development Fund FEDER, through the grant FIS2017-82855-P (MINECO/FEDER,UE) is acknowledged.



\section{Bibliography}
\begin{thebibliography}{45}
	\providecommand{\natexlab}[1]{#1}
	\providecommand{\url}[1]{\texttt{#1}}
	\expandafter\ifx\csname urlstyle\endcsname\relax
	\providecommand{\doi}[1]{doi: #1}\else
	\providecommand{\doi}{doi: \begingroup \urlstyle{rm}\Url}\fi
	
	\bibitem[Evans et~al.(1993)Evans, Cohen, and Morriss]{evans1993probability}
	Denis~J. Evans, E.~G.~D. Cohen, and G.~P. Morriss.
	\newblock Probability of second law violations in shearing steady states.
	\newblock \emph{Physical Review Letters}, 71\penalty0 (15):\penalty0
	2401--2404, oct 1993.
	\newblock \doi{10.1103/physrevlett.71.2401}.
	
	\bibitem[Evans and Searles(1994)]{evans1994equilibrium}
	Denis~J. Evans and Debra~J. Searles.
	\newblock Equilibrium microstates which generate second law violating steady
	states.
	\newblock \emph{Physical Review E}, 50\penalty0 (2):\penalty0 1645--1648, aug
	1994.
	\newblock \doi{10.1103/physreve.50.1645}.
	
	\bibitem[Gallavotti and
	Cohen(1995{\natexlab{a}})]{gallavotti1995dynamicalensembles}
	G.~Gallavotti and E.~G.~D. Cohen.
	\newblock Dynamical ensembles in nonequilibrium statistical mechanics.
	\newblock \emph{Physical Review Letters}, 74\penalty0 (14):\penalty0
	2694--2697, apr 1995{\natexlab{a}}.
	\newblock \doi{10.1103/physrevlett.74.2694}.
	
	\bibitem[Gallavotti and Cohen(1995{\natexlab{b}})]{gallavotti1995dynamical}
	G.~Gallavotti and E.~G.~D. Cohen.
	\newblock Dynamical ensembles in stationary states.
	\newblock \emph{Journal of Statistical Physics}, 80\penalty0 (5-6):\penalty0
	931--970, sep 1995{\natexlab{b}}.
	\newblock \doi{10.1007/bf02179860}.
	
	\bibitem[Jarzynski(1997{\natexlab{a}})]{jarzynski1997nonequilibrium}
	Christopher Jarzynski.
	\newblock Nonequilibrium equality for free energy differences.
	\newblock \emph{Physical Review Letters}, 78\penalty0 (14):\penalty0 2690,
	1997{\natexlab{a}}.
	
	\bibitem[Jarzynski(1997{\natexlab{b}})]{jarzynski1997equilibrium}
	C.~Jarzynski.
	\newblock Equilibrium free-energy differences from nonequilibrium measurements:
	A master-equation approach.
	\newblock \emph{Physical Review E}, 56\penalty0 (5):\penalty0 5018--5035, nov
	1997{\natexlab{b}}.
	\newblock \doi{10.1103/physreve.56.5018}.
	
	\bibitem[Crooks(1999)]{crooks1999a}
	Gavin~E. Crooks.
	\newblock Entropy production fluctuation theorem and the nonequilibrium work
	relation for free energy differences.
	\newblock \emph{Phys. Rev. E}, 60\penalty0 (3):\penalty0 2721--2726, Sep 1999.
	\newblock \doi{10.1103/PhysRevE.60.2721}.
	
	\bibitem[Jarzynski(2007)]{jarzynski2007comparison}
	Christopher Jarzynski.
	\newblock Comparison of far-from-equilibrium work relations.
	\newblock \emph{Comptes Rendus Physique}, 8\penalty0 (5):\penalty0 495--506,
	2007.
	
	\bibitem[Jarzynski(2011)]{jarzynski2011equalities}
	Christopher Jarzynski.
	\newblock Equalities and inequalities: Irreversibility and the second law of
	thermodynamics at the nanoscale.
	\newblock \emph{Annual Review of Condensed Matter Physics}, 2\penalty0
	(1):\penalty0 329--351, mar 2011.
	\newblock \doi{10.1146/annurev-conmatphys-062910-140506}.
	
	\bibitem[Esposito et~al.(2009)Esposito, Harbola, and
	Mukamel]{esposito2009nonequilibrium}
	Massimiliano Esposito, Upendra Harbola, and Shaul Mukamel.
	\newblock Nonequilibrium fluctuations, fluctuation theorems, and counting
	statistics in quantum systems.
	\newblock \emph{Reviews of modern physics}, 81\penalty0 (4):\penalty0 1665,
	2009.
	
	\bibitem[Campisi et~al.(2011)Campisi, H\"anggi, and Talkner]{campisi20111}
	Michele Campisi, Peter H\"anggi, and Peter Talkner.
	\newblock \textit{Colloquium}: Quantum fluctuation relations: Foundations and
	applications.
	\newblock \emph{Rev. Mod. Phys.}, 83:\penalty0 771--791, Jul 2011.
	\newblock \doi{10.1103/RevModPhys.83.771}.
	\newblock URL \url{http://link.aps.org/doi/10.1103/RevModPhys.83.771}.
	
	\bibitem[Seifert(2012)]{seifert2012stochastic}
	Udo Seifert.
	\newblock Stochastic thermodynamics, fluctuation theorems and molecular
	machines.
	\newblock \emph{Reports on Progress in Physics}, 75\penalty0 (12):\penalty0
	126001, 2012.
	\newblock URL \url{http://stacks.iop.org/0034-4885/75/i=12/a=126001}.
	
	\bibitem[Kurchan(2000)]{kurchan2000aquantum}
	Jorge Kurchan.
	\newblock A quantum fluctuation theorem, 2000.
	\newblock URL \url{https://arxiv.org/abs/cond-mat/0007360}.
	
	\bibitem[Perarnau-Llobet et~al.(2017)Perarnau-Llobet, Bäumer, Hovhannisyan,
	Huber, and Acin]{acin2017nogo}
	Mart{\'{\i}} Perarnau-Llobet, Elisa Bäumer, Karen~V. Hovhannisyan, Marcus
	Huber, and Antonio Acin.
	\newblock No-go theorem for the characterization of work fluctuations in
	coherent quantum systems.
	\newblock \emph{Physical Review Letters}, 118\penalty0 (7):\penalty0 070601,
	feb 2017.
	\newblock \doi{10.1103/physrevlett.118.070601}.
	
	\bibitem[Braginsky and Khalili(2003)]{braginsky2003quantum}
	Vladimir~B. Braginsky and Farid Y.~A. Khalili.
	\newblock \emph{Quantum Measurement}.
	\newblock Cambridge University Press, 2003.
	\newblock ISBN 052141928X.
	\newblock URL
	\url{https://www.ebook.de/de/product/3598942/vladimir_b_braginsky_farid_ya_khalili_kip_s_thorne_quantum_measurement.html}.
	
	\bibitem[Watanabe et~al.(2014)Watanabe, Venkatesh, and
	Talkner]{watanabe2014generalized}
	Gentaro Watanabe, B.~Prasanna Venkatesh, and Peter Talkner.
	\newblock Generalized energy measurements and modified transient quantum
	fluctuation theorems.
	\newblock \emph{Physical Review E}, 89\penalty0 (5), may 2014.
	\newblock \doi{10.1103/physreve.89.052116}.
	
	\bibitem[Chiara et~al.(2015)Chiara, Roncaglia, and Paz]{chiara2015measuring}
	Gabriele~De Chiara, Augusto~J Roncaglia, and Juan~Pablo Paz.
	\newblock Measuring work and heat in ultracold quantum gases.
	\newblock \emph{New Journal of Physics}, 17\penalty0 (3):\penalty0 035004, mar
	2015.
	\newblock \doi{10.1088/1367-2630/17/3/035004}.
	
	\bibitem[Deffner et~al.(2016)Deffner, Paz, and Zurek]{deffner2016quantum}
	Sebastian Deffner, Juan~Pablo Paz, and Wojciech~H. Zurek.
	\newblock Quantum work and the thermodynamic cost of quantum measurements.
	\newblock \emph{Physical Review E}, 94\penalty0 (1), jul 2016.
	\newblock \doi{10.1103/physreve.94.010103}.
	
	\bibitem[Allahverdyan and Nieuwenhuizen(2005)]{allahverdyan2005fluctuations}
	A.~E. Allahverdyan and Th.~M. Nieuwenhuizen.
	\newblock Fluctuations of work from quantum subensembles: The case against
	quantum work-fluctuation theorems.
	\newblock \emph{Physical Review E}, 71\penalty0 (6), jun 2005.
	\newblock \doi{10.1103/physreve.71.066102}.
	
	\bibitem[Campisi(2013)]{campisi2013quantum}
	Michele Campisi.
	\newblock Quantum fluctuation relations for ensembles of wave functions.
	\newblock \emph{New Journal of Physics}, 15\penalty0 (11):\penalty0 115008, nov
	2013.
	\newblock \doi{10.1088/1367-2630/15/11/115008}.
	
	\bibitem[Sone et~al.(2020)Sone, Liu, and Cappellaro]{sone2020quantum}
	Akira Sone, Yi-Xiang Liu, and Paola Cappellaro.
	\newblock Quantum jarzynski equality in open quantum systems from the one-time
	measurement scheme.
	\newblock \emph{Physical Review Letters}, 125\penalty0 (6):\penalty0 060602,
	aug 2020.
	\newblock \doi{10.1103/physrevlett.125.060602}.
	
	\bibitem[Roncaglia et~al.(2014)Roncaglia, Cerisola, and Paz]{roncaglia2014work}
	Augusto~J. Roncaglia, Federico Cerisola, and Juan~Pablo Paz.
	\newblock Work measurement as a generalized quantum measurement.
	\newblock \emph{Physical Review Letters}, 113\penalty0 (25), dec 2014.
	\newblock \doi{10.1103/physrevlett.113.250601}.
	
	\bibitem[Gherardini et~al.(2021)Gherardini, Belenchia, Paternostro, and
	Trombettoni]{gherardini2021endpoint}
	S.~Gherardini, A.~Belenchia, M.~Paternostro, and A.~Trombettoni.
	\newblock End-point measurement approach to assess quantum coherence in energy
	fluctuations.
	\newblock \emph{Phys. Rev. A}, 104:\penalty0 L050203, Nov 2021.
	\newblock \doi{10.1103/PhysRevA.104.L050203}.
	\newblock URL \url{https://link.aps.org/doi/10.1103/PhysRevA.104.L050203}.
	
	\bibitem[Chiara et~al.(2018)Chiara, Solinas, Cerisola, and
	Roncaglia]{chiara2018ancilla}
	Gabriele~De Chiara, Paolo Solinas, Federico Cerisola, and Augusto~J. Roncaglia.
	\newblock Ancilla-assisted measurement of quantum work.
	\newblock In \emph{Fundamental Theories of Physics}, pages 337--362. Springer
	International Publishing, 2018.
	\newblock \doi{10.1007/978-3-319-99046-0_14}.
	
	\bibitem[Chiara and Imparato(2022)]{chiara2022entropy}
	Gabriele~De Chiara and Alberto Imparato.
	\newblock Quantum fluctuation theorem for dissipative processes.
	\newblock \emph{Physical Review Research}, 4\penalty0 (2):\penalty0 023230, jun
	2022.
	\newblock \doi{10.1103/physrevresearch.4.023230}.
	
	\bibitem[Paule(2018)]{manzano2018thermodynamics}
	G.M. Paule.
	\newblock \emph{Thermodynamics and Synchronization in Open Quantum Systems}.
	\newblock Springer Theses. Springer International Publishing, 2018.
	\newblock ISBN 9783319939636.
	\newblock URL \url{https://books.google.es/books?id=ZLMatwEACAAJ}.
	
	\bibitem[Talkner et~al.(2007)Talkner, Lutz, and H\"anggi]{talkner2007c}
	Peter Talkner, Eric Lutz, and Peter H\"anggi.
	\newblock Fluctuation theorems: Work is not an observable.
	\newblock \emph{Physical Review E}, 75:\penalty0 050102, 2007.
	
	\bibitem[Tasaki(2000)]{tasaki2000jarzynski}
	Hal Tasaki.
	\newblock Jarzynski relations for quantum systems and some applications.
	\newblock \emph{arXiv}, cond-mat/0009244v2, 2000.
	
	\bibitem[Diosi(2011)]{diosi2011ashort}
	Lajos Diosi.
	\newblock \emph{A Short Course in Quantum Information Theory}.
	\newblock Springer Berlin Heidelberg, 2011.
	\newblock \doi{10.1007/978-3-642-16117-9}.
	
	\bibitem[Wiseman and Milburn(2009)]{wiseman2009quantum}
	Howard~M Wiseman and Gerard~J Milburn.
	\newblock \emph{Quantum measurement and control}.
	\newblock Cambridge university press, 2009.
	
	\bibitem[Albash et~al.(2013)Albash, Lidar, Marvian, and
	Zanardi]{albash2013fluctuation}
	Tameem Albash, Daniel~A. Lidar, Milad Marvian, and Paolo Zanardi.
	\newblock Fluctuation theorems for quantum processes.
	\newblock \emph{Physical Review E}, 88\penalty0 (3):\penalty0 032146, sep 2013.
	\newblock \doi{10.1103/physreve.88.032146}.
	
	\bibitem[Diosi(1988)]{diosi1988cqm}
	L.~Diosi.
	\newblock {Continuous quantum measurement and Ito formalism}.
	\newblock \emph{Phys. Lett. A}, 129\penalty0 (8-9):\penalty0 419--423, 1988.
	
	\bibitem[Braunstein and Caves(1988)]{braunstein1988quantum}
	Samuel~L. Braunstein and Carlton~M. Caves.
	\newblock Quantum rules: an effect can have more than one operation.
	\newblock \emph{Foundations of Physics Letters}, 1\penalty0 (1):\penalty0
	3--12, mar 1988.
	\newblock \doi{10.1007/bf00661312}.
	
	\bibitem[Breuer and Petruccione(2002)]{breuer2002theory}
	Heinz-Peter Breuer and Francesco Petruccione.
	\newblock \emph{The theory of open quantum systems}.
	\newblock Oxford University Press on Demand, 2002.
	
	\bibitem[Sokolovski et~al.(2019)Sokolovski, Brouard, and
	Alonso]{sokolovski2019fromquantum}
	D~Sokolovski, S~Brouard, and D~Alonso.
	\newblock From quantum to classical by numbers.
	\newblock \emph{New Journal of Physics}, 21\penalty0 (12):\penalty0 123031, dec
	2019.
	\newblock \doi{10.1088/1367-2630/ab59b7}.
	\newblock URL \url{https://doi.org/10.1088%2F1367-2630%2Fab59b7}.
	
	\bibitem[Deffner et~al.(2010)Deffner, Abah, and Lutz]{deffner2010quantumwork}
	Sebastian Deffner, Obinna Abah, and Eric Lutz.
	\newblock Quantum work statistics of linear and nonlinear parametric
	oscillators.
	\newblock \emph{Chemical Physics}, 375\penalty0 (2-3):\penalty0 200--208, oct
	2010.
	\newblock \doi{10.1016/j.chemphys.2010.04.042}.
	
	\bibitem[Nielsen and Chuang(2010)]{nielsen2010quantum}
	Michael~A Nielsen and Isaac~L Chuang.
	\newblock \emph{Quantum computation and quantum information}.
	\newblock Cambridge university press, 2010.
	
	\bibitem[Streltsov et~al.(2017)Streltsov, Adesso, and
	Plenio]{streltsov2017quantum}
	Alexander Streltsov, Gerardo Adesso, and Martin~B. Plenio.
	colloquium: Quantum coherence as a resource.
	\newblock \emph{Reviews of Modern Physics}, 89\penalty0 (4), oct 2017.
	\newblock \doi{10.1103/revmodphys.89.041003}.
	
	\bibitem[Groenewold(1971)]{groenewold1971aproblem}
	H.~J. Groenewold.
	\newblock A problem of information gain by quantal measurements.
	\newblock \emph{International Journal of Theoretical Physics}, 4\penalty0
	(5):\penalty0 327--338, sep 1971.
	\newblock \doi{10.1007/bf00815357}.
	
	\bibitem[Santos et~al.(2019)Santos, Céleri, Landi, and
	Paternostro]{santos2019therole}
	Jader~P. Santos, Lucas~C. Céleri, Gabriel~T. Landi, and Mauro Paternostro.
	\newblock The role of quantum coherence in non-equilibrium entropy production.
	\newblock \emph{npj Quantum Information}, 5\penalty0 (1):\penalty0 23, March
	2019.
	\newblock ISSN 2056-6387.
	\newblock URL \url{https://doi.org/10.1038/s41534-019-0138-y}.
	
	\bibitem[Baumgratz et~al.(2014)Baumgratz, Cramer, and
	Plenio]{baumgratz2014quantifying}
	T.~Baumgratz, M.~Cramer, and M.~B. Plenio.
	\newblock Quantifying coherence.
	\newblock \emph{Phys. Rev. Lett.}, 113:\penalty0 140401, Sep 2014.
	\newblock \doi{10.1103/PhysRevLett.113.140401}.
	\newblock URL \url{https://link.aps.org/doi/10.1103/PhysRevLett.113.140401}.
	
	\bibitem[Witten(2020)]{witten2018anintroduction}
	Edward Witten.
	\newblock A mini-introduction to information theory.
	\newblock \emph{La Rivista del Nuovo Cimento}, 43\penalty0 (4):\penalty0
	187--227, mar 2020.
	\newblock \doi{10.1007/s40766-020-00004-5}.
	
	\bibitem[Gaspard(2013)]{gaspard2013entropy}
	Pierre Gaspard.
	\newblock Entropy production in the quantum measurement of continuous
	observables.
	\newblock \emph{Physics Letters A}, 377\penalty0 (3):\penalty0 181 -- 184,
	2013.
	\newblock ISSN 0375-9601.
	\newblock \doi{https://doi.org/10.1016/j.physleta.2012.11.036}.
	\newblock URL
	\url{http://www.sciencedirect.com/science/article/pii/S0375960112012157}.
	
	\bibitem[Jensen(1906)]{jensen1906surlesfonctions}
	J.~L. W.~V. Jensen.
	\newblock Sur les fonctions convexes et les in{\'{e}}galit{\'{e}}s entre les
	valeurs moyennes.
	\newblock \emph{Acta Mathematica}, 30\penalty0 (0):\penalty0 175--193, 1906.
	\newblock \doi{10.1007/bf02418571}.
	
	\bibitem[Naghiloo et~al.(2020)Naghiloo, Tan, Harrington, Alonso, Lutz, Romito,
	and Murch]{naghiloo2020heat}
	M.~Naghiloo, D.~Tan, P.~M. Harrington, J.~J. Alonso, E.~Lutz, A.~Romito, and
	K.~W. Murch.
	\newblock Heat and work along individual trajectories of a quantum bit.
	\newblock \emph{Phys. Rev. Lett.}, 124:\penalty0 110604, Mar 2020.
	\newblock \doi{10.1103/PhysRevLett.124.110604}.
	\newblock URL \url{https://link.aps.org/doi/10.1103/PhysRevLett.124.110604}.
	
\end{thebibliography}
\end{document}